\documentclass[10pt,conference]{IEEEtran}
\IEEEoverridecommandlockouts
\usepackage{cite}
\usepackage{amsmath,amssymb,amsfonts}
\usepackage{algorithmic}
\usepackage{graphicx}
\usepackage{textcomp}
\usepackage{xcolor,colortbl}
\usepackage{booktabs}
\usepackage{mathtools}
\usepackage[english]{babel}
\usepackage{multirow}
\usepackage{subfig}
\usepackage{url}
\usepackage{hyperref}
\usepackage{mathabx}
\usepackage{amsmath}
\usepackage{amsthm}
\usepackage{amssymb}
\usepackage{stmaryrd}
\usepackage[export]{adjustbox}
\usepackage[english,ruled,vlined,linesnumbered]{algorithm2e}
\usepackage[justification=centering]{caption}

\theoremstyle{definition}
\newtheorem{property}{Property}

\newtheorem{definition}{Definition}

{\renewcommand{\arraystretch}{1.5}}

\newcommand{\comment}[1]{}
\newcommand{\colored}{\cellcolor{gray!50}}

\def\BibTeX{{\rm B\kern-.05em{\sc i\kern-.025em b}\kern-.08em
    T\kern-.1667em\lower.7ex\hbox{E}\kern-.125emX}}

\AtBeginDocument{
  \catcode`_=12
  \begingroup\lccode`~=`_
  \lowercase{\endgroup\let~}\sb
  \mathcode`_="8000
}

\newcommand{\Prob}{\mbox{Pr}}

\begin{document}

\title{Mining Java Memory Errors using Subjective Interesting Subgroups with Hierarchical Targets}

\author{Anonymous}


\author{\IEEEauthorblockN{Youcef Remil\IEEEauthorrefmark{1}\IEEEauthorrefmark{2},
Anes Bendimerad\IEEEauthorrefmark{2},
Mathieu Chambard\IEEEauthorrefmark{3}, 
Romain Mathonat\IEEEauthorrefmark{2}, 
Marc Plantevit\IEEEauthorrefmark{4}
and Mehdi Kaytoue\IEEEauthorrefmark{1}\IEEEauthorrefmark{2}}
\IEEEauthorblockA{\IEEEauthorrefmark{1}Univ Lyon, INSA Lyon, CNRS, LIRIS UMR 5205, F-69621, Lyon, France}
\IEEEauthorblockA{\IEEEauthorrefmark{2}Infologic, 99 avenue de Lyon, F-26500 Bourg-L{\`{e}}s-Valence, France}
\IEEEauthorblockA{\IEEEauthorrefmark{3}Ecole Nationale Supérieure de Rennes, FR-35170, Bruz, France}
\IEEEauthorblockA{\IEEEauthorrefmark{4}EPITA Research  Laboratory (LRE), FR-94276, Le Kremlin-Bicetre, France}

}

\maketitle

\begin{abstract}

Software applications, particularly Enterprise Resource Planning (ERP) systems, are crucial to the day-to-day operations of many industries, where it is essential to maintain these systems effectively and reliably. In response, Artificial Intelligence for Operations Systems (AIOps) has emerged as a dynamic framework, harnessing cutting-edge analytical technologies like machine learning and big data to enhance incident management procedures by detecting, predicting, and resolving issues while pinpointing their root causes. In this paper, we leverage a promising data-driven strategy, dubbed Subgroup Discovery (SD), a data mining method that can automatically mine incident datasets and extract discriminant patterns to identify the root causes. However, current SD solutions have limitations in handling complex target concepts with multiple attributes organized hierarchically. We illustrate this scenario by examining the case of Java out-of-memory incidents among several possible applications. We have a dataset that describes these incidents, including their topology and the types of Java objects occupying memory when it reaches saturation, with these types arranged hierarchically. This scenario inspires us to propose a novel Subgroup Discovery approach that can handle complex target concepts with hierarchies. To achieve this, we design a new Subgroup Discovery framework along with a pattern syntax and a quality measure that ensure the identified subgroups are relevant, non-redundant, and resilient to noise. To achieve the desired quality measure, we use the Subjective Interestingness model that incorporates prior knowledge about the data and promotes patterns that are both informative and surprising relative to that knowledge. We apply this framework to investigate out-of-memory errors and demonstrate its usefulness in root-cause diagnosis. Notably, this paper stands out as a contribution to both Data Mining and Java memory analysis research. To validate the effectiveness of our approach and the quality of the retrieved patterns, we present an empirical study conducted on a real-world scenario from our ERP system.

\end{abstract}

\section{Introduction}\label{sec:introduction}

\begin{figure*}[t]
\centering
\includegraphics[width=0.95\textwidth]{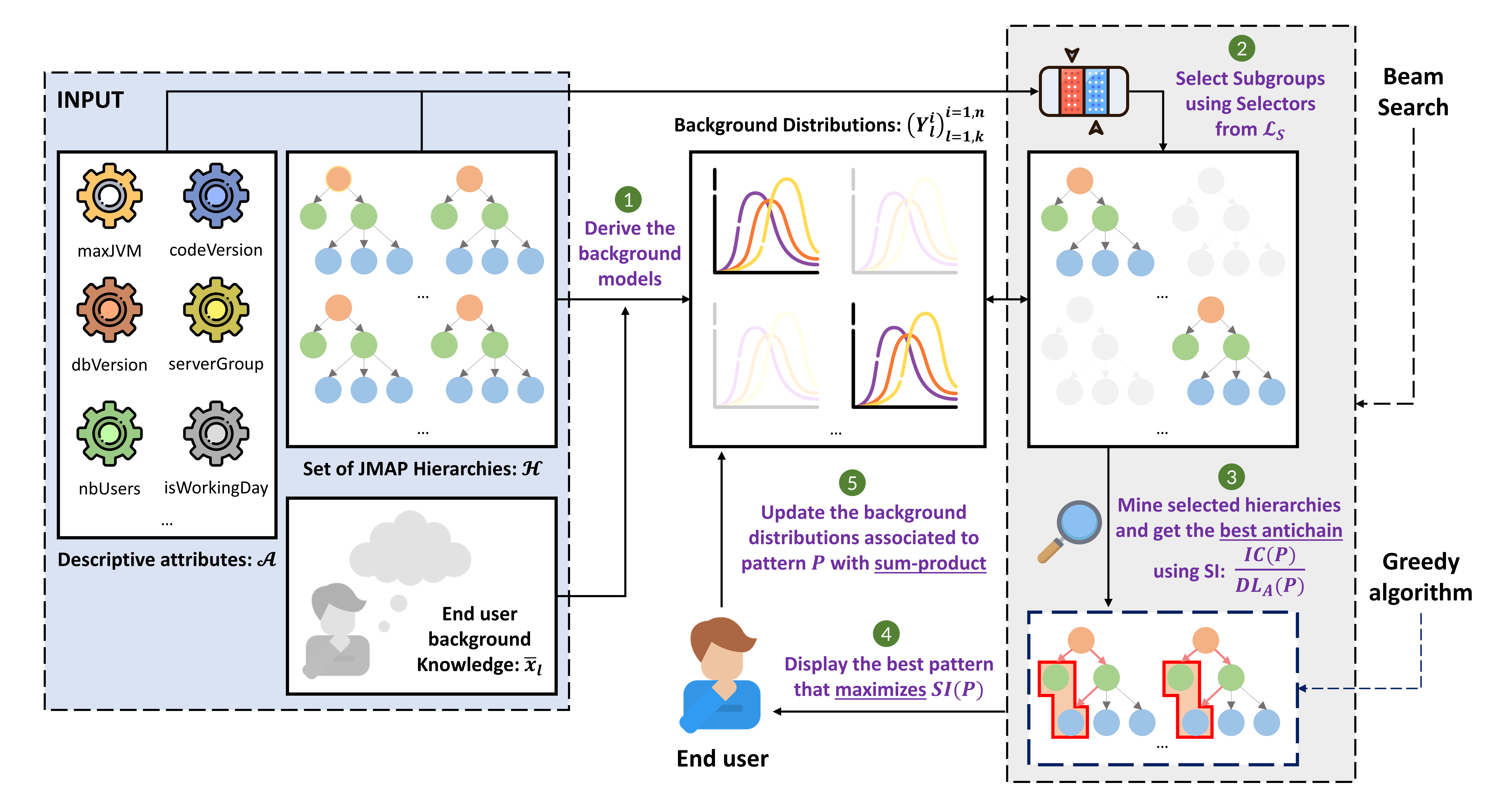}
\caption{\label{fig:overview} Overview of our Subjectively Interesting Subgroup Discovery Framework with hierarchical target concepts.}
\end{figure*}

\begin{figure}[t]
\centering
\begin{tabular}{cc}
\includegraphics[width=0.45\textwidth]{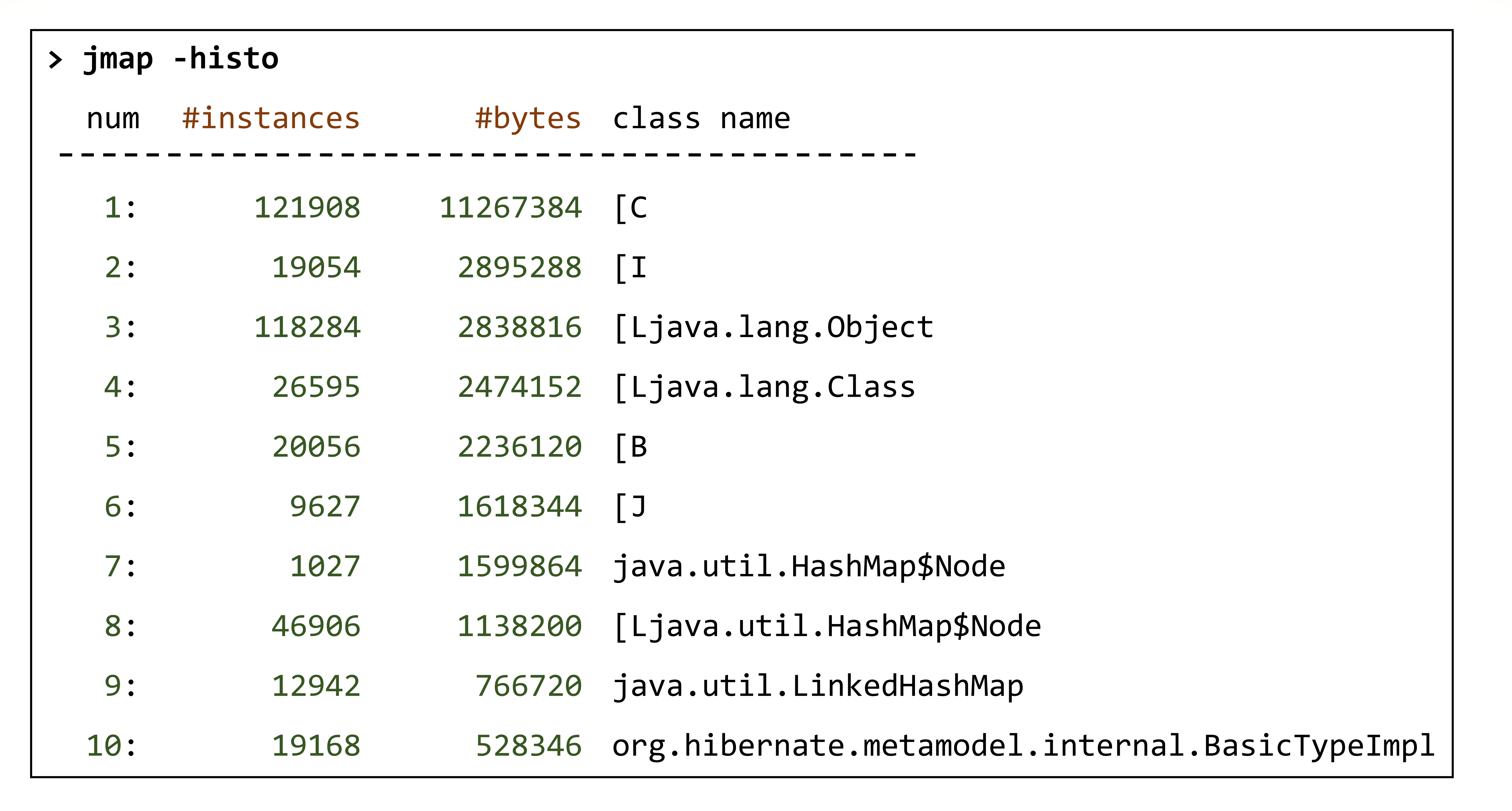}
\end{tabular}
\caption{\label{fig:jmap} Example of a \texttt{jmap} histogram.}
\end{figure}

Industries are rapidly moving toward digitization thanks to the notable progress made in AI and software development. At the heart of this revolution, Enterprise Resource Planning (ERP) software systems are directly connected to factories and their equipments. They provide a large panel of tools and data to users across different services of a company and significantly help them to achieve their daily tasks. With the increasing importance of these software systems in numerous industry activities, there is a growing need to ensure their optimal performance. Collecting data from various sources within the ERP ecosystem allows organizations to gain valuable insights into operational efficiency and identify areas for improvement. In this context, Artificial Intelligence for IT Operations (AIOps) emerges as a game-changing technology that refers to the application of advanced data-driven algorithms including machine learning and big data analytics to intelligently improve, strengthen, and automate a wide range of IT operations~\cite{dang2019aiops,bogatinovski2021artificial,becker2020towards,notaro2021survey}. Therefore, organizations are turning to AIOps in order to detect, prevent, diagnose, and rapidly mitigate high-impact incidents~\cite{chen2020towards}. One of the most common incidents encountered is related to \texttt{OutOfMemory} errors, which occur when the memory allocated to the software is fully utilized, often due to a memory leak caused by some specific bug. Engineers usually rely on tools that provide statistics about the memory content at the time of the error, which aids in identifying the root cause. For example, the \texttt{jmap} command depicts the memory consumed by each class in a Java Virtual Machine, as shown in Fig.~\ref{fig:jmap}. Moreover, these classes are  organized \textit{hierarchically} in packages. For example, the class \texttt{LinkedHashMap} in Line~$9$ is part of the package \texttt{java.util}, which is itself included in the package \texttt{java}. By analyzing the histogram, analysts can identify classes that consume significantly more memory than expected, which may be the underlying root-cause of memory saturation. 

Analyzing histograms can be an overwhelming task for several reasons. First, the analyst may lack a clear understanding of what constitutes ``normal" size consumption for each class, requiring significant experience or reference histograms with a ``healthy'' memory consumption. In Fig.~\ref{fig:jmap}, for example, the \texttt{[C} class may not be the cause of the incident despite being the most consuming class, as this value may be its usual size. Secondly, some memory incidents are not related to a single class but to a software feature that impacts several classes belonging to specific packages. This raises the question of how to concisely identify the suspicious packages and/or classes without redundancy, especially when dealing with many hierarchy levels. 
Finally, analysts often have to inspect a vast dataset of histograms related to various incidents described by their contexts (e.g., software version and type) as well as \texttt{jmap} histograms describing memory content.

In this case,  the analyst seeks to find contexts that significantly co-occur with specific kinds of memory errors to pinpoint the root cause for several incidents at once. Existing methods~\cite{DBLP:conf/sigsoft/XieA05,DBLP:conf/icse/JungLRP14,DBLP:conf/wosp/WeningerGM19,DBLP:journals/jss/SorS14} primarily focus on diagnosing Java memory problems separately, without analyzing large sets of incidents simultaneously to discover common patterns. Also, most of them address particularly the problem of memory leak detection, whereas there are other possible factors that can cause \texttt{OutOfMemory} incidents. Our goal is to address these challenges, by providing a generic approach that extracts useful information from such datasets.

Mining datasets to identify subgroups that exhibit some property of interest is known as Subgroup Discovery (SD)~\cite{DBLP:books/mit/fayyadPSU96/Klosgen96,DBLP:journals/widm/Atzmueller15}. Among many possibilities, SD aims to find patterns describing subsets of data where the distribution of a target variable deviates from the ``norm''. However, none of existing SD approaches have considered the case when the target concept consists in multiple attributes organized as hierarchies, such as \texttt{jmap} histograms. To address this limitation, we introduce a novel and generic SD setting that elegantly handles this case. We employ Subjective Interestingness framework~\cite{DBLP:journals/datamine/Bie11} to model the information dependency between hierarchical attributes, and assess informativeness of patterns. This framework has been successfully instantiated in several data mining tasks~\cite{DBLP:journals/datamine/KapoorSL21,DBLP:journals/datamine/DengKLB21,DBLP:journals/datamine/BendimeradMLPRB20}. Its strength lies in its ability to include the user's prior knowledge about the data to find unexpected patterns. It also allows to iteratively update the interestingness model to account for information already transmitted to the user during the mining task. We characterize subgroups with specific subsets of target attributes called \textit{antichains} (set of hierarchically incomparable elements). This antichain constraint allows us to avoid redundancy within the same pattern, as hierarchically related attributes often transmit the same information. Different from the method  proposed in~\cite{DBLP:conf/kdd/BendimeradLPRB19} which extracts contrastive attributes but from a single hierarchy, our solution mines a dataset of many hierarchies described by additional contextual attributes. Our solution is then applied to analyze a dataset of \texttt{jmap} histograms and contextualize memory errors, although it remains generic and suitable for numerous SD tasks with hierarchical target attributes.

Figure~\ref{fig:overview} provides an overview of the entire process underlying our proposed approach. The input dataset comprises incident data, including \texttt{jmap} histograms and other relevant attributes that contextualize each incident. Additionally, the approach leverages prior information pertaining to the dataset under analysis. As an example, these priors could consist of \texttt{jmap} histograms computed from "healthy" servers, serving as benchmarks for expected class sizes. Our approach utilizes these priors to establish background distributions that represent "healthy" memory usage patterns. Subsequently, it sifts through the incident dataset to identify the most informative subgroups, pinpointing suspicious classes and packages that exhibit significant deviations from the established background distribution. Once this subgroup is identified, it is communicated to the analyst. Following this, the background model is updated to assimilate this newfound information, which the analyst is already aware of. This iterative process can be repeated as necessary, allowing for a refined understanding of the dataset's anomalies and patterns. Each of the framework phases will be discussed in detail in the upcoming sections.

\noindent \textbf{Contributions.} Our contribution is three-fold: 
(1) Unlike existing Java memory analysis methods that address individual memory issues within specific use cases, we present a comprehensive approach that tackles a wide range of related incidents stemming from various root causes, which allows for a more holistic diagnosis. (2) We present a pioneering approach to address a novel Subgroup Discovery challenge where the target concept consists of multiple attributes organized in a hierarchy by introducing an adapted pattern syntax and a new interestingness measure rooted in Subjective Interestingness, which properly guides the search for informative and non-redundant patterns, along with a simple yet effective beam-search algorithm to search for interesting subgroups. (3) Employing both qualitative and quantitative analysis, we showcase the effectiveness of our approach by utilizing a real-world ERP dataset to mine contextualized \texttt{jmap} histograms, thereby diagnosing the root causes of memory errors.

\section{Methodology}~\label{sec:methodology}

We performed our analysis on a dataset comprising approximately 4,000 Java memory snapshots, also known as \texttt{Java memory heap dumps}, which were collected from over 350 servers over a three-month period. Each memory snapshot represents the instantiated objects residing in the Java virtual machine heap of a specific server at a specific moment. Each snapshot is associated with a hierarchical structure that defines a directed acyclic dependency graph, grouping classes and/or sub-packages within the same package, along with information about the size of their instantiated objects. Furthermore, we enriched these snapshots with additional descriptive features, including environmental variables, resulting in a wealth of properties that provide contextual information for Subgroup Discovery.

\subsection{Raw Data}
\begin{figure}[t]
\centering
\begin{tabular}{cc}
\includegraphics[width=0.44\textwidth]{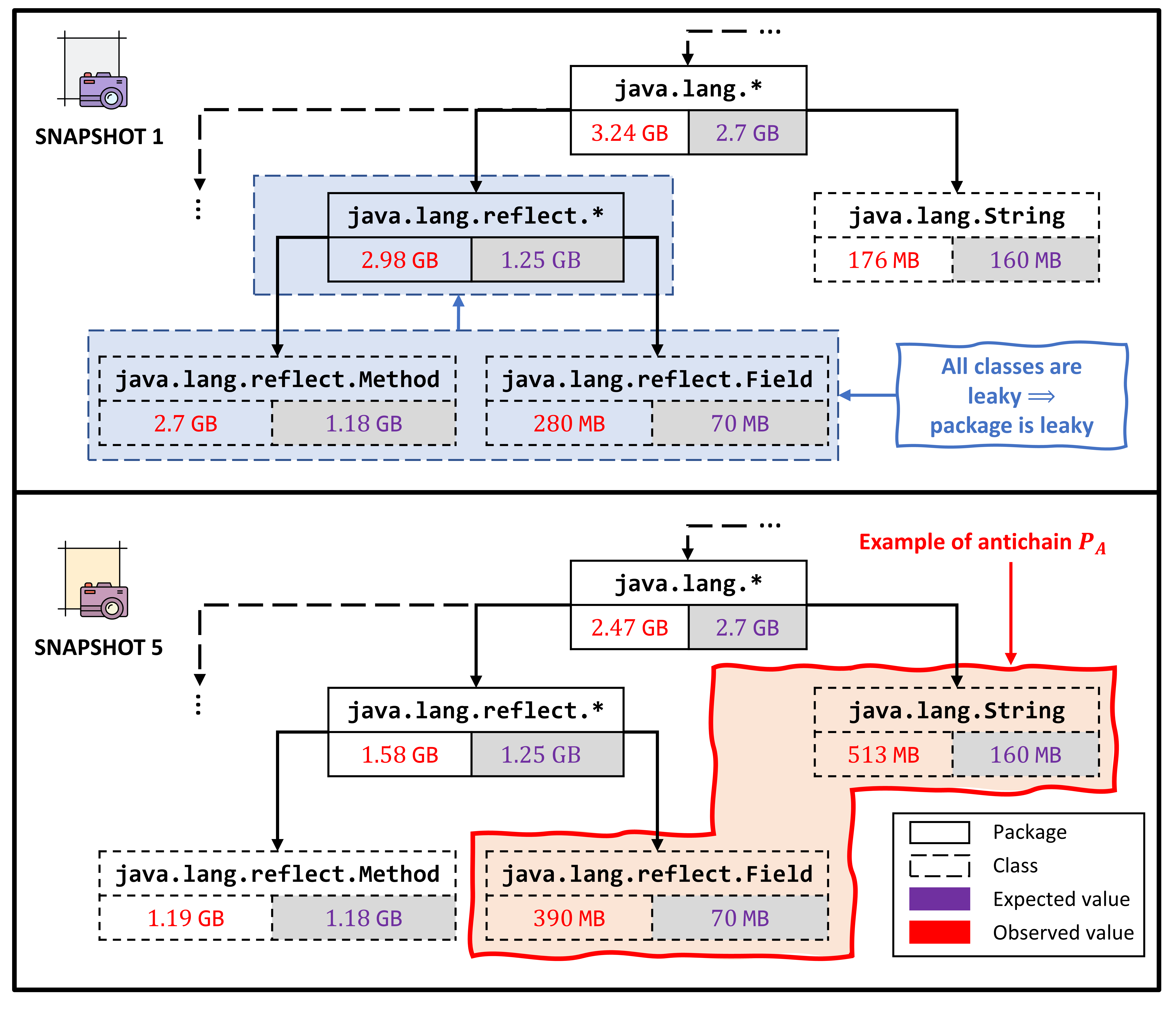}
\end{tabular}
\caption{\label{fig:exampleHierarchy} Example of \texttt{JMAP} sub-hierarchy retrieved from a specific memory snapshot with some class/package sizes.}
\end{figure}

\smallbreak
\noindent \textbf{Java Memory Heap Dumps.} The analysis of Java memory plays a crucial role in monitoring the performance of Java applications. Java virtual machine (\texttt{JVM}) uses the heap memory to store objects created by an application, and garbage collection is performed to reclaim memory resources that are no longer used. However, the \texttt{JVM} may not be able to free up memory if objects still retain references, which can lead to memory saturation and cause the \texttt{java.lang.OutOfMemoryError} exception to be thrown. This error can be due to insufficient allocated heap size, a coding error that causes memory leaks, or loading a large number of objects into memory simultaneously. To obtain specific memory-related statistics, we leverage the command-line utility \texttt{jmap} with the \texttt{-histo} option. This allows us to obtain a class-wise histogram that displays the number of instantiated objects in the heap per class, the total memory used for each of these objects, and the fully qualified class name. We use this histogram for each memory snapshot to build a hierarchy that groups the classes and/or sub-packages recursively into their parent packages. The sub-hierarchy illustrated in Fig.~\ref{fig:exampleHierarchy} demonstrates that the package size is the total size of all its sub-packages or classes. For instance, the \texttt{java.lang.} package contains the sub-package \texttt{java.lang.reflect.} as well as the class \texttt{java.lang.String}.

\smallbreak
\noindent \textbf{Topology information for contextualization.} Each memory snapshot is accompanied by additional descriptive characteristics. Particularly, these features describe the execution environment and the software component in which the exception \texttt{OutOfMemoryError} is triggered. For instance, our ERP software is characterized by its application identifier, 
its software version, its declination among main server families such as sales, factory, etc. The JVM can be parameterized by the flag \texttt{Xmx} which specifies the maximum memory allocation pool for the \texttt{JVM}, the \texttt{Xms} to indicate the initial memory allocation pool. Information on when the memory collapsed is also provided (e.g., whether it is a working day).

\subsection{A Unified Data Model}

\begin{table*}
    \centering
     \caption{Toy Example of a dataset $(\mathcal{O},\mathcal{A},\mathcal{H})$. Gray cells indicates that the size values are larger than what was expected.}
    \renewcommand{\arraystretch}{1.1}
    \scalebox{0.85}{\begin{tabular}{@{}c|cccc|ccccc}
\toprule
\multirow{2}{*}{$\mathcal{O}$} &
  \multicolumn{4}{c|}{\textbf{Descriptive attributes $\mathcal{A}$}} &
  \multicolumn{5}{c}{\begin{tabular}[c]{@{}c@{}}\textbf{Size of instantiated objects w.r.t. packages in MB}\\ represented with hierarchies $\mathcal{H}$ (see example for snapshots $o_1$ and $o_5$ in Fig.~\ref{fig:exampleHierarchy})\end{tabular}} \\* \cmidrule(l){2-10} 
 &
  \multicolumn{1}{c|}{softType} &
  \multicolumn{1}{c|}{softVersion} &
  \multicolumn{1}{c|}{\texttt{Xmx}} &
  weekDay &
  \multicolumn{1}{c|}{\texttt{J.L.*}} &
  \multicolumn{1}{c|}{\texttt{J.L.reflect.*}} &
  \multicolumn{1}{c|}{\texttt{J.L.reflect.Field}} &
  \multicolumn{1}{c|}{\texttt{J.L.reflect.Method}} &
  \texttt{J.L.String} \\ \hline 
$o_1$ &
  \multicolumn{1}{c|}{Sales} &
  \multicolumn{1}{c|}{V\_$3$} &
  \multicolumn{1}{c|}{$4.2e+09$} &
  True &
  \multicolumn{1}{c|}{$3242$} &
  \multicolumn{1}{c|}{\colored$2980$} &
  \multicolumn{1}{c|}{\colored$280$} &
  \multicolumn{1}{c|}{\colored$2700$} &
  $176$ \\ \hline
$o_2$ &
  \multicolumn{1}{c|}{Sales} &
  \multicolumn{1}{c|}{V\_$3$} &
  \multicolumn{1}{c|}{$2.3e+09$} &
  False &
  \multicolumn{1}{c|}{$3296$} &
  \multicolumn{1}{c|}{\colored$3003$} &
  \multicolumn{1}{c|}{\colored$322$} &
  \multicolumn{1}{c|}{\colored$2678$} &
  \colored$355$ \\ \hline
$o_3$ &
  \multicolumn{1}{c|}{EDI} &
  \multicolumn{1}{c|}{V\_$1$} &
  \multicolumn{1}{c|}{$6.4e+09$} &
  True &
  \multicolumn{1}{c|}{$2305$} &
  \multicolumn{1}{c|}{$1474$} &
  \multicolumn{1}{c|}{\colored$264$} &
  \multicolumn{1}{c|}{$1210$} &
  $163$ \\ \hline
$o_4$ &
  \multicolumn{1}{c|}{Factory} &
  \multicolumn{1}{c|}{V\_$1$} &
  \multicolumn{1}{c|}{$1.8e+09$} &
  False &
  \multicolumn{1}{c|}{$2217$} &
  \multicolumn{1}{c|}{$1481$} &
  \multicolumn{1}{c|}{\colored$386$} &
  \multicolumn{1}{c|}{$1095$} &
  \colored$480$ \\ \hline
$o_5$ &
  \multicolumn{1}{c|}{Factory} &
  \multicolumn{1}{c|}{V\_$2$} &
  \multicolumn{1}{c|}{$2.4e+09$} &
  True &
  \multicolumn{1}{c|}{$2475$} &
  \multicolumn{1}{c|}{$1582$} &
  \multicolumn{1}{c|}{\colored$390$} &
  \multicolumn{1}{c|}{$1192$} &
  \colored$513$ \\ \hline
$o_6$ &
  \multicolumn{1}{c|}{Manager} &
  \multicolumn{1}{c|}{V\_$2$} &
  \multicolumn{1}{c|}{$5.3e+09$} &
  True &
  \multicolumn{1}{c|}{$2016$} &
  \multicolumn{1}{c|}{$1258$} &
  \multicolumn{1}{c|}{$56$} &
  \multicolumn{1}{c|}{$1202$} &
  $140$ \\ \hline
$o_7$ &
  \multicolumn{1}{c|}{Sales} &
  \multicolumn{1}{c|}{V\_$3$} &
  \multicolumn{1}{c|}{$2.4e+09$} &
  True &
  \multicolumn{1}{c|}{$3398$} &
  \multicolumn{1}{c|}{\colored$2814$} &
  \multicolumn{1}{c|}{\colored$320$} &
  \multicolumn{1}{c|}{\colored$2494$} &
  \colored$402$ \\ \hline
$o_8$ &
  \multicolumn{1}{c|}{Factory} &
  \multicolumn{1}{c|}{V\_$3$} &
  \multicolumn{1}{c|}{$8.2e+09$} &
  False &
  \multicolumn{1}{c|}{$2715$} &
  \multicolumn{1}{c|}{$1326$} &
  \multicolumn{1}{c|}{$84$} &
  \multicolumn{1}{c|}{$1200$} &
  $147$ \\ \hline
$o_9$ &
  \multicolumn{1}{c|}{Sales} &
  \multicolumn{1}{c|}{V\_$3$} &
  \multicolumn{1}{c|}{$6.4e+09$} &
  True &
  \multicolumn{1}{c|}{$2430$} &
  \multicolumn{1}{c|}{$1577$} &
  \multicolumn{1}{c|}{\colored$412$} &
  \multicolumn{1}{c|}{$1165$} &
  $120$ \\ \hline
$o_{10}$ &
  \multicolumn{1}{c|}{Sales} &
  \multicolumn{1}{c|}{V\_$1$} &
  \multicolumn{1}{c|}{$4.5e+09$} &
  True &
  \multicolumn{1}{c|}{$2570$} &
  \multicolumn{1}{c|}{$1283$} &
  \multicolumn{1}{c|}{$68$} &
  \multicolumn{1}{c|}{$1215$} &
  \colored$422$ \\ \hline
\end{tabular}
}
 
  \label{tab:toyDataset}
\end{table*}

We create a dataset $\mathcal{D} = (\mathcal{O},\mathcal{A},\mathcal{H})$ that unifies the different data sources mentioned above. $\mathcal{O}=\{o_i\}_{1\leq i \leq n}$ is a set of objects that correspond to memory snapshots indexed by the pairs \texttt{(server;timestamp)}. $\mathcal{A}=(a_j)_{1 \leq j \leq m}$ is a vector of descriptive attributes used to contextualize the snapshots, and $\mathcal{H}=\{H_i\}_{1 \leq i \leq n}$ is a set of hierarchies constructed from \texttt{jmap} histograms. These hierarchies group classes and/or sub-packages that belong to the same package and include the size of their instantiated Java objects.

Each attribute $a$ can be represented as a mapping $a : \mathcal{O} \longrightarrow R_a$, where $R_a$ is called the domain of attribute $a$. $R_a$ is given by $\mathbb{R}$ if $a$ is numerical, by a finite set of categories $C_i$ if $a$ is categorical, or by $\{0, 1\}$ if $a$ is Boolean. Table~\ref{tab:toyDataset} provides a dataset with $10$ objects $\mathcal{O}=\{o_1,...,o_{10}\}$ corresponding to $10$ memory snapshots generated during \texttt{outOfMemoryError} exceptions. Each object is described by $4$ attributes and referenced by a hierarchy $H \in \{H_1,...,H_{10}\}$. For example, the attribute \texttt{softType} is nominal and has $4$ possible values. The attribute \texttt{Xmx} is numerical, and the attribute \texttt{weekDay} is a Boolean that indicates whether the memory crash occurred on a working day.  Fig.~\ref{fig:exampleHierarchy} illustrates with a typical example of sub-hierarchical elements that have varying values of both $H_1$ and $H_5$, which highlight the integer-valued attributes associated with class sizes and their corresponding packages. For instance, the package \texttt{java.lang.reflect} in $H_1$ has a size of $2.98$ GB and contains only two classes, namely, \texttt{java.lang.reflect.Method} (size of $2.7$ GB) and \texttt{java.lang.reflect.Field} (size of $280$ MB).   

\subsection{Hierarchical Target Concepts}

We consider the scenario where the concepts of interest are defined as a set of positive integer-based attributes that are structured hierarchically. These hierarchically organized concepts are generally referred to as \textit{counters} based on their observed values. These counters represent either the size of the classes (which are located at the leaves of the hierarchy) or packages (which are located at internal nodes). The root node represents the size of the heap at the time of the memory crash. We formally define a hierarchy $H \in \mathcal{H}$ as follows.

\begin{definition}[Hierarchy]
A hierarchy $H_i \in \mathcal{H}$ is defined as a tuple $H_i = (E^{(i)},\leq, \langle e_1, x^{(i)}_{1}\rangle)$ for $i \in \llbracket 1, n \rrbracket$ where:
\begin{itemize}
\item $E^{(i)}=\{\langle e_1, x^{(i)}_{1}\rangle,...,\langle e_k, x^{(i)}_{k}\rangle\}$ is a set of $k$ items (nodes or concepts) with their counters. For convenience, We use sometimes $E = \{e_1,..., e_k\}$ to refer to the set of items without their counters. 
\item $\leq$ is a partial order relation defined over this set $E$, indicating the relationship of predecessors between hierarchically linked concepts,
\item $\forall e \in E: e_1 \leq e$ (the item $e_1$ is called the root of $\mathcal{H}$) 
\item there is only one path from the root $e_1$ to any other item:
\[ \forall e_j,e_k,e_l \in E: e_j \leq e_l \wedge e_k \leq e_l \Longrightarrow e_j \leq e_k \vee e_k \leq e_j .\]
\end{itemize}
\end{definition}

For example, in Fig.~\ref{fig:exampleHierarchy}, both of the following relations hold: $(\text{\texttt{java.lang.*}}) \leq  (\text{\texttt{java.lang.reflect}})$ and $(\text{\texttt{java.lang.*}}) \leq  (\text{\texttt{java.lang.reflect.Field}})$. 

Moreover, we assume that the counter value $x_l$ of a concept $e_l$ is always larger or equal than the value of its successors. We introduce the following operations used throughout the remainder of this paper. Given $S \subseteq E$, we have: 
\begin{itemize}
\item Precedessors operator $\Uparrow$, and successors operator $\Downarrow$ as:
{\small
\begin{align*}
\Uparrow S&=\{ e \in E \mid \exists e' \in S: e \leq e'\},\\
\Downarrow S&=\{ e \in E \mid \exists e' \in S: e' \leq e\},
\end{align*}}

\item Strict predecessor relation: $e_j < e_k \Leftrightarrow e_j \leq e_k \wedge e_j \neq e_k, $

\item The direct successor relation $\prec$ as: $e_j \prec e_k \Longleftrightarrow \Downarrow \{e_j\} \cap \Uparrow \{e_k\} = \{ e_j,e_k\}$. Also, if $e_j \prec e_k $, we use the notation $\pi_k=j$ to refer to the index of the only direct parent of $e_k$ ($e_j=e_{\pi_k}$),

\end{itemize}

The counter value of a concept $e_{l}$ for an object $o_i \in \mathcal{O}$ 
is denoted using the discrete random variable $X_{l}^{(i)} \in \mathbb{N}$. If a particular value of a concept $e_{l}$ is empirically observed for the object $o_i \in \mathcal{O}$, it is denoted in the hierarchy $H_i$ by $\hat{x}^{(i)}_l$.   

\subsection{Contrastive Antichains as patterns}\label{sec:ca}
We first introduce the concept of contrastive antichains as interesting patterns based on a single hierarchy. These patterns are comprised of a subset of hierarchically disjoint concepts that are informative and non-redundant. In our case study, we aim to concisely inform developers  about suspicious classes and packages. To evaluate the interestingness of a pattern, we rely on prior knowledge about counters, such as developers' rough estimates of the space occupied by classes in the heap. For example, in Fig.~\ref{fig:exampleHierarchy}, developers expect the size of the \texttt{java.lang.string} class to be around $160$ MB based on the analysis conducted on healthy servers. Thus, discovering that this class takes up $513$ MB in snapshot $o_5$ is surprising.

Providing the user with such concepts can be interesting. However, because one counter's information affects the expectation of other hierarchically related concepts, one intuition is to recursively aggregate the interesting concepts at the same level into a higher level of the hierarchy. This approach is shown in Fig.\ref{fig:exampleHierarchy}, where, for example, instead of listing both the leaking classes \texttt{java.lang.reflect.Method} and \texttt{java.lang.reflect.Field}, we only provide their parent package \texttt{java.lang.reflect}. Another intuition is to only include non-comparable concepts in the pattern (i.e., no concept is a predecessor or successor of any other concept in the same pattern). We refer to this set of concepts by an \textit{antichain}. Thus, an antichain denoted as $P_A$ ensures that the information provided is not redundant. For each identified concept $e_l \in P_A$ for a specific object $o_i \in \mathcal{O}$, a contrastive antichain pattern informs the user about the value $\hat{x}^{(i)}_l$. Yet, users generally tend to memorize only the order of magnitude indication instead of precise values. Hence, we refer to the counters in patterns with the scale $\lfloor log_{2}(\hat{x}^{(i)}_l) \rfloor$. 

\begin{definition}[Contrastive Antichains]
Given a hierarchy $H_i = (E^{(i)},\leq, \langle e_1, x^{(i)}_{1}\rangle)$ with pairs of concepts and their values $\langle e_l, x^{(i)}_{l}\rangle \in E^{(i)}$, a contrastive antichain pattern  $P_A \subseteq E$ is a subset of concepts that form an antichain w.r.t. $\leq$, i.e., $\forall e_j, e_k \in P_A$: $e_j \leq e_k \Longrightarrow e_j=e_k$, with the integers $\lfloor \log(\hat{x}^{(i)}_{l}) \rfloor$ describing the scale of the values of their counters.
\end{definition}


\subsection{Need to Characterize Subgroups with a Common Antichain}
Developers often analyze a large set of objects (memory snapshots) at once. Providing contrastive antichains for each individual object can be overwhelming, and many objects may share hierarchical concepts that have similar properties. We need to simultaneously characterize a subset of objects that are together associated to a contrastive antichain pinpointing suspicious classes or packages.
An interesting example in Table~\ref{tab:toyDataset} is the subgroup $\{o_2, o_4, o_5, o_7\}$ containing memory snapshots from virtual machines with a \texttt{Xmx} flag value not exceeding $2.5e+09$. All these snapshots exhibit unexpectedly high memory consumption of classes forming the following antichain: \{\texttt{java.lang.reflect.Field, java.lang.String}\}. Identifying interesting subgroups along with their antichains necessitates a unified \textit{pattern language}. 
Since there is an extremely large number of subgroups that can be derived from combinations of descriptive attributes and target antichains, an automatic Subgroup Discovery approach can prove invaluable by exploring a set of candidate hypotheses and using an quality function to score subgroups and identify the best of them.
We exploit the subjective interestingness framework (SI) proposed in~\cite{DBLP:journals/datamine/Bie11}. This framework makes it possible to iteratively incorporate the new information provided to the user when communicating a subgroup to her, to avoid communicating subgroups with redundant information. For instance, suppose the user is presented with the subgroup $\{o_1, o_2, o_7, o_9\}$ defined by Sales servers of Version $3$ associated with an antichain consisting of only the package \texttt{java.lang.reflect}. Subsequently, the user is presented with another subgroup containing objects $\{o_2, o_4, o_5, o_7\}$, associated with the antichain \{\texttt{java.lang.reflect.Field, java.lang.String}\}. Although the second pattern is interesting, it becomes less surprising when the user is aware of the first pattern, because $\{o_2, o_7\}$ are already associated with the concept \texttt{java.lang.reflect}, which is a package that already includes  \texttt{java.lang.reflect.field}. Hence, our approach should ignore this pattern and suggest a more restrictive one, such as the subgroup that covers only $\{o_4, o_5\}$ with the description $(\text{\texttt{Xmx}} < 2.5e+09 \wedge \text{softType} = \text{Factory})$.

\section{Subjectively Interesting Subgroups with Contrastive Antichains in Hierarchies}~\label{sec:method}

To efficiently exploit Subgroup Discovery with hierarchical target concepts, we need to address many challenges: (1) handling the complex data structure including both subgroup descriptions and antichains, (2) assessing pattern interestingness related to a specific subgroup and its retrieved antichain, (3) defining a mining algorithm that is scalable and can identify subgroups and antichains that maximize the proposed measure of interestingness, (4) implementing an effective mechanism to incrementally update the user's background knowledge.

\subsection{Pattern Language}~\label{subsec:pl}
We consider a pattern language defined as a pair $\mathcal{L} = (\mathcal{L}_S,\mathcal{L}_A)$ such that $\mathcal{L}_S$ is the \textit{subgroup pattern language} defined over descriptive attributes $\mathcal{A}$, and $\mathcal{L}_A$ is the \textit{antichain pattern language} defined over the concepts $E$ from $\mathcal{H}$. A pattern $P \in \mathcal{L}$ is then given as $P = (P_s, P_A)$, where $P_s \in \mathcal{L}_S$ is a constrained selector of a subset of objects using their descriptive attribute values and $P_A \in \mathcal{L}_A$ is a retrieved antichain from $E$. In more details, the subgroup pattern language is defined as $\mathcal{L}_S = \bigtimes_{j=1}^{\mid \mathcal{A} \mid} {Sel}_j$, where ${Sel}_j$ is a selector applied over an attribute $a_j \in \mathcal{A}$ that describes a set of objects based on their attribute values (e.g., it is given by the set of all possible intervals in $\mathbb{R}$ if $a_j$ is numerical). Hence, $P_s \in \mathcal{L}_S$ is then given by a set of restrictions over each descriptive attribute i.e., $P_s = (Sel_j)_{1 \leq j \leq m}$. These patterns are ordered from the most general to the most restrictive by an order relation $\sqsubseteq$. More precisely, for two patterns $P_s = ({Sel}_j)_{1 \leq j \leq m} \in \mathcal{L}_S$ and ${P'}_s = ({Sel'}_j)_{1 \leq j \leq m} \in \mathcal{L}_S$, we have: $P_s \sqsubseteq {P'}_s \Leftrightarrow \forall j \in \llbracket 1, m \rrbracket \, ({Sel}_j \supseteq {Sel'}_j)$. On the other hand, the antichain pattern language is defined as the set of all possible antichains that can be derived from $E$ i.e., $\mathcal{L}_A = \{ P_A \subseteq E \mid \forall e_l, e_k \in P_A$: $e_l \leq e_k \Longrightarrow e_l=e_k \}$.

\smallbreak
\noindent\textbf{Subgroup patterns and objects.} A subgroup pattern $P_s = ({Sel}_j)_{1 \leq j \leq m}$ is referred to as \textit{covering} an object $o \in \mathcal{O}$ iff $\forall j \in \llbracket 1, m \rrbracket : a_j(o) \in {Sel}_j$. A subgroup pattern $P_s$ \textit{covers} a set $O \subseteq \mathcal{O}$ iff it covers each object $o \in \mathcal{O}$. Using the \textit{cover} concept, we define the function $\delta(O) \in \mathcal{L}_S$ which gives the most restrictive subgroup pattern that covers a set of objects $O$: $\forall P_s \in \mathcal{L}_S$, $P_s$ covers $O$ iff $P_s \sqsubseteq \delta(O)$. For a given subgroup pattern $P_s$, the set of all objects covered by $P_s$ is referenced by the \textit{extent} concept: $ext(P_s)=\{o \in \mathcal{O} \mid P_s \sqsubseteq \delta(o)\} $ \cite{DBLP:conf/iccs/GanterK01}. 

\begin{definition}[Subgroup]
A subgroup is any subset of objects $s \in \mathcal{O}$ that can be selected using a pattern $P_s$ over descriptive attributes $\mathcal{A}$. The set of all possible subgroups is denoted $\mathcal{S}=ext(\mathcal{L}_S)=\{ext(P_s) \mid P_s \in \mathcal{L}_S\}$. In other terms, a subgroup is a set of objects that can be characterized with some restrictions of attributes, turning it interpretable to the user. 

\end{definition}

\subsection{Subjective Interestingness Measure}~\label{subsec:si}

\vspace{-3mm}

We design a subjective interestingness measure to assess the quality of each pattern $P = (P_s, P_A)$. This function measures its surprisingness when contrasted with some background distribution that represents user priors about the data. Therefore, we need to formally model the prior beliefs about each counter $\langle e_l, x^{(i)}_{l}\rangle \in E^{(i)}$. We will present these counters through the probability distributions $Pr (X^{(i)}_{l} = x^{(i)}_{l})$.

\smallbreak
\noindent \textbf{Background Distributions.} For each concept in the hierarchy, we assume that we have a reference that derives either an approximation of its value or its proportion to other concepts that are hierarchically dependent on it. In our case, we compute \texttt{jmap} histograms on normally behaving servers to derive the expected values of the size or size proportion of each class and/or package in the memory heap. Our goal is to represent the expected value $x_l$ of each concept $e_l \in E$, as well as the expectations conditioned on the parent except for the root, i.e., $x_l \mid x_{\pi_l}$. As an example, the \texttt{java.lang.String} class is expected to have an average size of approximately $160$MB, and the package \texttt{java.lang.reflect} is expected to contain about $95$\% of the \texttt{java.lang.reflect.Method} class.

Given these two constraints, there are an infinite number of possible solutions for defining probability distributions for all concepts. To address this, we follow the approach of~\cite{DBLP:journals/datamine/Bie11} and consider only the distributions that maximize entropy, meaning that they do not introduce additional assumptions beyond the explicitly specified expectations that would reduce the entropy. As noted in~\cite{DBLP:conf/kdd/BendimeradLPRB19}, this approach results in geometric probability distributions for each random variable $X^{(i)}_{l}$, hence: 

{\small
\begin{property} \label{prop:marginal}
The marginal distribution for each random variable $X^{(i)}_l$ is geometric, and it is given as:
$$
\Prob(X^{(i)}_l = x_l)=\left(1-\frac{1}{1+\bar{x}_l}\right)^{x_l} \cdot \frac{1}{1+\bar{x}_l}.
$$
\end{property}}

\smallbreak
\noindent \textbf{Interestingness of a Pattern.} Now that a probability distribution has been defined to model the user's background knowledge, we need to evaluate the antichain $P_A$ on each object of the subgroup $o \in ext(P_s)$ with respect to its probability distribution. This is necessary to efficiently determine the most interesting and surprising patterns that contradict the background beliefs or the previously discovered findings (i.e., after iteratively updating its primary knowledge). To assess the score of a given pattern $P = (P_s, P_A) \in \mathcal{L}$, we propose a new quality measure rooted in the framework of subjective interestingness SI~\cite{DBLP:journals/datamine/Bie11}. $SI(P)$ is defined as the ratio between the information content of the pattern $P$ and its description length: $SI(P)=\frac{IC(P)}{DL(P)}$.

The \textit{Information Content} (IC) measures the amount of information communicated to the user. It is defined as the negative log probability under the background distribution: $IC(P)=-\log(\Prob(P))$. However, calculating the information content of such complex pattern can be challenging, especially when dealing with multiple probability distributions of the same concept associated with different objects in the subgroup. To address this, we aim to quantify the information gain provided by an effective aggregation of counters $\langle e_l, \hat{x}^{(i)}_{l}\rangle$ in $P_A$, by evaluating the objects of the subgroup under their respective probability distributions (which might have been updated in previous iterations). A straightforward solution would be to use the mean value of each concept in the subgroup. However, the mean is too sensitive to outliers and does not provide a complete overview of the subgroup values w.r.t. a concept, as no assumptions are made about the subgroup variance.

To overcome this issue, we came out with the idea of calculating the cumulative distribution function  from the quantile $q_{\alpha}$ of order $\alpha$ that indicates the value below which a certain percentage of the subgroup values fall, for each concept. The hyperparameter ($0 < \alpha < 1$) is selected based on the quality of the subgroups returned. For instance, if the quantile of order $0.25$ is considered, then our solution informs the user that $75$\% of the subgroup values are greater than this quantile. To better understand this approach, let's consider the subgroup $s = \{o_1, o_2, o_7, o_9\}$ with corresponding values $\{2980, 3003, 2814, 1577\}$ for the package \texttt{java.lang.reflect}. Assuming that all objects have the same geometric probability distribution with a mean of $1250$MB (in the first iteration), we can use the cumulative distribution function under the quantile of order $0.25$ to calculate the probability $\Prob(X^{(i)}_{\texttt{java.lang.reflect.}} \geq 2814) = 0.08$, which is interesting. In other words, for the subgroup defined as $(\text{softType} = \text{Sales} \wedge \text{softVersion} = \text{V\_}3)$, $75\%$ of its objects have a size that is greater than or equal to $2814$MB for the \texttt{java.lang.reflect.} package. This information is valuable and surprisings since it deviates from what we would expect based on its probability distribution.

As previously stated in Sec.~\ref{sec:ca}, the information content is communicated to the user by transmitting the scales of the values, instead of the exact values.  As a consequence, we define the new random variables $Y^{(i)}_l = \lfloor \log_{2}(\hat{X}^{(i)}_{l}) \rfloor$. Concretely, the $IC$ of a pattern $P \in \mathcal{L}$ is given as follows:   

{\small
\begin{align*}
IC(P) &=   IC \left( (P_s, P_A) \right) = -\log \left( \prod_{o_i \in s} \prod_{e_l \in P_A} \Prob(Y^{(i)}_l \geq {q_{\alpha}}^{s}_l) \right), 
\\ &= - \sum_{o_i \in s} \sum_{e_l \in P_A} \log \left(\Prob(Y^{(i)}_l \geq {q_{\alpha}}^{s}_l) \right), 
\\ &= - \sum_{o_i \in s} \sum_{e_l \in P_A} \log \left( (1-p_l)^{2^{{q_{\alpha}}^{s}_l}} - (1-p_l)^{2^{{q_{\alpha}}^{s}_l + 1}} \right).
\end{align*}}
where ${q_{\alpha}}^{s}_l$ is the quantile of order $\alpha$ of the values $\{y^{(i)}_{l}  \}_{o_i \in s}$ for the concept $e_l \in P_A$

\smallbreak
The \textit{description length} (DL) is a measure of the complexity involved in communicating a pattern $P$ to a user. In our case we propose to compute it based on both the subgroup pattern $P_s$ and the antichain $P_A$. When communicating the antichain, items closer to the root $e_1$ are more likely to be familiar to the user and easier to interpret. For instance, it is simpler to communicate the package \texttt{java.lang} ($2^{\text{nd}}$ level) than the class \texttt{java.lang.reflect.Method} ($4^{\text{th}}$ level). On the other hand, a subgroup with only a few selectors is easier to interpret as it helps to quickly pinpoint the root cause. In order to characterize as many objects as possible in a subgroup that is distinguished by an interesting antichain, we avoid linear penalization of the subgroup size. Hence, the DL is given as:
{\small
\begin{align*}
& DL(P) = DL_s(P_s) \cdot DL_A(P_A)
\\ &= \left( \beta \cdot \log(|s|) + \gamma \cdot \Vert P_s \Vert \right) \cdot \left( \eta \cdot (\sum_{e_l \in P_A} 1 + \log(|\Uparrow \{ e_l\}|)) \right)
\end{align*}}
with $\beta$, $\gamma$ and $\eta$ are hyperparameters to weight each part of the DL according to the user preferences.  

\subsection{Updating the Background Knowledge}~\label{subsec:update}
When conveying a pattern to the user, her background knowledge model must be updated to take into consideration the new piece of information. The communicated pattern values are likely to become the new expected values, and therefore, the probability distributions must also be updated accordingly.  In the following, let $\mathcal{R} \subseteq \mathcal{L}$ be the set of patterns that have already been observed up to the $\text{i}^\text{th}$ iteration, that is, $\mathcal{R} = {(P^{1}_s, P^{1}_A),...,(P^{(i)}_s, P^{(i)}_A)}$. We refer to the quality of the pattern $P$ assuming the user has knowledge of $\mathcal{R}$ as:

$
SI(P \mid \mathcal{R}) =\frac{IC(P \mid \mathcal{R})}{DL(P)}=\frac{-\log(\Prob(P \mid \mathcal{R}))}{DL(P)}
$

The probability $\Prob(P \mid \mathcal{R})$ represents the likelihood of pattern $P$ appearing in the data given that the user is aware of the quantile of order $\alpha$ for the subgroup values of each concept for all previously communicated patterns $P' = (P_s', P_A') \in \mathcal{R}$. In other words, instead of knowing the exact value of each object, the user only knows that its probability being higher than ${q_{\alpha}}^{s'}_l$ is $(1-\alpha)$. Thus, we update the probability distribution $\Prob(Y^{(i)}_l = y_l)$ as follows: 

{\small
\begin{align*}
  \Prob(Y^{(i)}_l = y_l)=\begin{cases}
    \Prob(Y^{(i)}_l = y_l) \cdot \frac{1-\alpha}{\Prob(Y^{(i)}_l \geq {q_{\alpha}}^{s'}_l)}, & \text{if $Y^{(i)}_l \geq {q_{\alpha}}^{s'}_l$}.\\
    \Prob(Y^{(i)}_l = y_l) \cdot \frac{\alpha}{1 - \Prob(Y^{(i)}_l \geq {q_{\alpha}}^{s'}_l)}, & \text{otherwise}.
  \end{cases}
\end{align*}}

For example, assuming that the user has been given the subgroup $s' = \{o_1, o_2, o_7, o_9\}$, whose antichain contains the package \texttt{java.lang.reflect} and that the first quartile (i.e., $\alpha = 0.25$) has been used to retrieve useful patterns, then the new probability distribution for each of the objects in $s'$ must verify:  $\Prob(Y^{(i)}_{java.lang.reflect} \geq \lfloor \log_{2}(2814) \rfloor) = 0.75$.

The hierarchical organization of the concepts implies that updating the probability distribution of a particular concept will have a direct and recursive impact on its predecessors and successors. For instance, if the user becomes certain (with a probability of 75\%) that the size of the \texttt{java.lang.reflect.} package is larger than 2814, then it follows that the size of its parent, \texttt{java.lang.}, must also be larger than 2814 with a probability greater than 75\%, since $X_{\pi_l} > X_l$. These dependencies between the random variables can be represented using a Bayesian tree, which is a graphical model. To propagate the impact of updating some random variables to all nodes in the hierarchy, we use the sum-product inference algorithm~\cite{DBLP:books/lib/Bishop07}.


\subsection{Mining Interesting Patterns}~\label{subsec:algo}
The process of finding the most interesting patterns in our context is very costly, since the computational complexity of a subgroup discovery task is known to be prohibitive and results from the huge size of the search space $|\mathcal{S}|$ that increases exponentially. Besides, each generated subgroup pattern $P_s \in \mathcal{L_s}$ has to be evaluated with the set of all possible antichains $L_A$. Moreover, the interestingness measure used in this paper SI is not monotonic (i.e., it is not straightforward to derive non-trivial bounds on its values to prune some uninteresting patterns), which implies that exhaustive search is not a feasible strategy to be adopted. We employ optimization procedures that are commonly used in both scenarios (i.e., enumeration of subgroup patterns and the search for contrastive antichains). We derive \texttt{SCA-Miner}, a heuristic approach that uses beam search to generate at each level of the lattice the most interesting subgroups with their associated antichain which is retrieved with a greedy search algorithm w.r.t. the subjective interestingness measure.

The proposed algorithm is outlined in Algorithm~\ref{algo:sca-miner} (\texttt{SCA-MINER}), which is an iterative approach, that aims at each iteration to provide the user with an interesting pattern $P$. The algorithm starts by updating the model using the sum-product method to incorporate the user's previously acquired knowledge $\mathcal{R}$. Then, it employs a beam search strategy to enumerate the subgroup patterns and use a greedy search algorithm to derive the associated antichain that maximizes the SI. The algorithm continues until the beam search yields an empty pattern or the maximum size of the best patterns collection $\mathcal{P}$ is reached. The beam search systematically explores the conjunctions of selectors by expanding a limited set of patterns that have the largest SI so far. It evaluates the subgroup patterns on their set of hierarchies to extract the best associated contrastive antichain. In this phase, a greedy search method is used to build an antichain for a specific subgroup pattern. This process is repeated on each level in the beam search, where only the most promising patterns are maintained. The mining process stops when all possible selectors are explored or a chosen stopping criterion is met (e.g., the search depth). The algorithm outputs the best pattern found throughout the search.    

\begin{algorithm}[t]
    \caption{\texttt{SCA-Miner}~\label{algo:sca-miner}}

  \KwIn{\footnotesize the dataset: $\mathcal{D}_{crash} = (\mathcal{O},\mathcal{A},\mathcal{H})$, ${width}$: the number of most promising subgroups per level,
  ${depth}$: the maximum depth to explore in the lattice,
  $threshold$: a threshold on the number of patterns.}
  \KwOut{\footnotesize $\mathcal{P}$: An ordered collection of patterns $P \in \mathcal{L}$ sorted based on iteratively updated $SI$.}
  {\footnotesize
  $\mathcal{P} \gets \langle \rangle$ \\
  $\mathcal{R} \gets \emptyset$ \\
    \Repeat{$P_s=\emptyset$ or $P_A=\emptyset$ or $|\mathcal{P}| = threshold$} {
    // Update the model with the sum-product algorithm \\
    // and derive the probabilities $\Prob(Y^{(i)}_l = \hat{y}^{(i)}_l \mid \mathcal{R}):$ \\
    Sum-product($\mathcal{H}$, $\mathcal{R}$) \\
    // Get the best subgroup along with its associated contrastive antichain: \\
	$(P_s, P_A) \gets $ \texttt{BeamSearch}($\mathcal{D}$, $width$, $depth$, $\mathcal{R}$)\\
	\If{ $P_s \neq \emptyset$ and $P_A \neq \emptyset$}{
		$\mathcal{P}.append ((P_s, P_A))$ \\
		$ \mathcal{R} \gets \mathcal{R} \cup (P_s, P_A)$ }
		}
	}
\end{algorithm}


\section{Experiments}~\label{sec:xp}
In this section, we present the experimental study we conducted to evaluate the quality of results provided by \texttt{SCA-Miner}~\footnote{\url{https://github.com/RemilYoucef/sca-miner}} to analyze Java memory errors reported by our ERP software. We aim to assess whether the approach is capable of identifying interesting patterns based on Subjective Interestingness and whether the update of the background model is effective. Additionally, we assess the interpretability and relevance of the identified patterns in pinpointing potential root causes regarding our case study. 

\subsection{Experimental Setup and Methodology}

\noindent \textbf{Datasets and hyperparameters.} Our experimental study involved analyzing more than 4,000 Java memory heap dumps collected over a 3-month period from approximately 350 servers. To establish reference values for the average heap space usage of each class or package, we generated a separate dataset of heap dumps from healthy servers at the beginning of each week. We only considered the top 200 classes in each histogram associated with a heap dump, as these are often the ones that retain the most heap space. The resulting dataset comprised 3,320 memory snapshots, each described by 14 descriptive attributes, and mapped into hierarchies to contextualize each subgroup and antichain. We used the readily available data mining tool \texttt{Pysubgroup}~\cite{DBLP:conf/pkdd/Lemmerich018} to extend the beam search algorithm to fit our pattern language and measure of interest. We set the beam width to 50 and the number of selective selectors to 4, and displayed the top 20 patterns based on Subjective Interestingness (SI). We set the quantile of order 20 to inform users about 80\% of the subgroup data values for a specific concept. The hyperparameters associated with the DL function were set empirically as $\beta = 0.8, \gamma = 0.2$, and $\eta = 1$. 


\smallbreak
\noindent \textbf{Baselines.} While there are no approaches in the literature that specifically support hierarchical Subgroup Discovery with interesting target concepts, we consider some baselines in our study to highlight the benefits of \texttt{SCA-MINER}'s novel features, including the hierarchical structure and the new interestingness measure, as well as its capability to iteratively update the user background knowledge to avoid redundancy. First, we compare with the \texttt{SI} approach, which returns the best results according to our interestingness measure, but does not iteratively update the background model. Next, we compare with \texttt{Customized WRAcc (CWRAcc)}, which adapts the widely-used WRAcc measure~\cite{DBLP:journals/jmlr/LavracKFT04} to our problem, measuring the deviation of the subgroup mean value from the mean value of the entire dataset regarding a target concept. Specifically, $CWRAcc (P_s, P_A, \theta) = \frac{1}{{|P_A|}^\theta} \sum_{e_l \in P_A}(\frac{1}{|s|} \sum_{o_i \in s}\hat{x}^{(i)}_l-\bar{x}_l)$. We run the Beam Search algorithm under the same hyperparameters as \texttt{SCA-MINER}. We also compare against the \texttt{KL} divergence, as performed in~\cite{DBLP:journals/datamine/LeeuwenK12}, to measure the difference between the observed probability distributions for the contextualized memory snapshots and the expected probability distribution, at each level of the hierarchy. Finally, we perform a post-processing step \texttt{PP} for each of these baselines to eliminate redundant patterns according to the Jaccard coefficient, as performed in the work of~\cite{DBLP:conf/flairs/AtzmullerP08}.  

\smallbreak
\noindent \textbf{Comparative Metrics:} To evaluate the effectiveness of \texttt{SCA-MINER} compared to baselines, we utilize two metrics: \textbf{Average Contrast} and \textbf{Redundancy}. The first metric measures the contrast between the observed subgroup values $\hat{x}^{(i)}_l$ and their expected values $\bar{x}_l$ for each retrieved pattern. We propose using the contrast metric defined as follows for a pattern $P$: $contrast(P)=\sum{e_l \in P_A}\frac{\frac{1}{|s|} (\sum{o_i \in s}\hat{x}^{(i)}l)-\bar{x}l}{\frac{1}{|s|} \sum{o_i \in s}\hat{x}^{(i)}l}$. The second metric, redundancy, measures the redundancy between patterns communicated to the user while accounting for the redundancy between hierarchically related concepts of different antichains with respect to the same subgroup elements. To accomplish this, we propose the following redundancy measure evaluated on a set of patterns: $redund(\mathcal{P})=\frac{\sum{P \in \mathcal{P}} max{P' \in \mathcal{P} \backslash {P}} Jaccard(s,s') \cdot \frac{| {e_l \in P_A \mid \exists e_{k} \in P_A': e_l \leq e_k \vee e_k \leq e_l}|}{|P_A \cup P_A'|}}{|\mathcal{P}|}$.

\subsection{Results interpretation}

\smallbreak
\noindent \textbf{Comparative evaluation.} Fig.~\ref{fig:metrics} displays the comparative measures used to evaluate \texttt{SCA-MINER} against the baseline techniques discussed earlier. The SI measure produces larger contrast values in general (with or without update) compared to the CWRAcc and KL-Divergence measures. This is due to the fact that CWRAcc typically retrieves antichains containing nodes with higher counters, which do not necessarily result in contrastive patterns, as their average contrast is remarkably low when compared to that of the SI measure. Nonetheless, the contrast values of the non-updated SI tend to be higher than those of the updated SI because the algorithm often focuses on the same top patterns, generating a high level of redundancy. This redundancy is demonstrated in the bar graphs, where more than $60\%$ of the patterns are redundant, which makes the process less surprising for the end user. Similarly, when using the CWRAcc and KL-Divergence measures, the resulting patterns often contain redundancy. Even after applying post-processing to the final pattern set, there is still a much higher level of redundancy when compared to using the updated SI measure (only about $4\%$ redundancy).  

\comment{
\begin{figure}[]
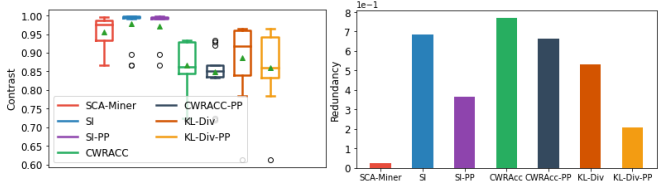

\centering
\begin{center}  
 \begin{tabular}{c|c}
 
\toprule
\textbf{Contrast} &  \textbf{Redundancy} \\
 \midrule
 \includegraphics[width=0.48\linewidth]{XP/contrast.png} &
  \includegraphics[width=0.48\linewidth]{XP/redund.png}\\
 \bottomrule
   \end{tabular}
\end{center}
\caption{\label{fig:metrics} Comparison of contrast and redundancy between top patterns of \texttt{SCA-MINER} against the baselines.}
\end{figure}}

\begin{figure}
\centering
\includegraphics[width=1.0\linewidth]{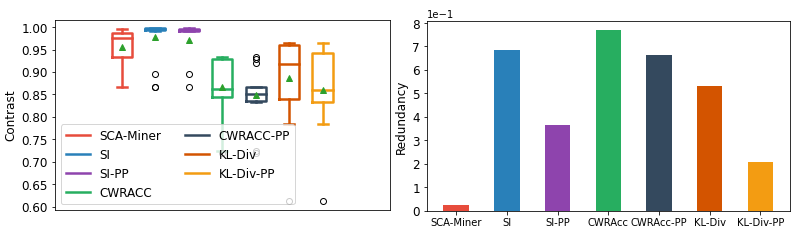} \\

\caption{\label{fig:metrics} Comparison of contrast and redundancy between top patterns of \texttt{SCA-MINER} against the baselines.}
\end{figure}

\begin{figure}
\centering
\includegraphics[width=0.85\linewidth]{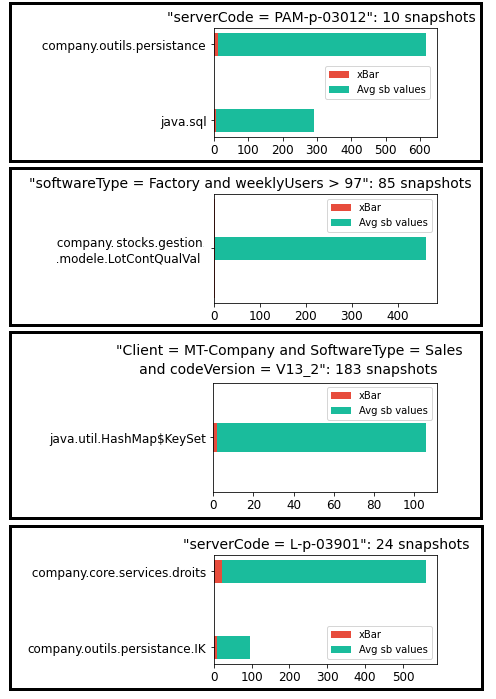} \\

\caption{\label{fig:toppatterns} Top patterns returned by \texttt{SCA-MINER}}
\end{figure}

\begin{table}
    \centering
     \caption{Statistics related to antichains that belong to the top patterns retrieved with \texttt{SCA-MINER}}
    \renewcommand{\arraystretch}{0.9}
    \scalebox{0.75}{\begin{tabular}{@{}ccccccc}
\toprule
\multicolumn{1}{c|}{\textbf{Top $k$}}                 & \multicolumn{1}{c|}{\textbf{Antichains}}                                                                            & \multicolumn{1}{c|}{\textbf{$\bar{x}$}} & \textbf{Min} & $q_{0.2}$ & \textbf{Avg} & \textbf{Max} \\ \midrule
\multicolumn{1}{c|}{\multirow{2}{*}{Top $1$}} & \multicolumn{1}{l|}{\texttt{company.outil.persistance}}                                                              & \multicolumn{1}{c|}{$13$}      & $568$ & $579$ & $606$ & $659$ \\
\multicolumn{1}{c|}{}                        & \multicolumn{1}{l|}{\texttt{java.sql}}                                                                                    & \multicolumn{1}{c|}{$7$}       & $267$ & $271$ & $284$ & $305$ \\ \midrule
\multicolumn{1}{c|}{Top $2$}                  & \multicolumn{1}{l|}{\begin{tabular}[l]{@{}l@{}}\texttt{company.stock.gestion}.\\ \texttt{modele.LotContQualVal}\end{tabular}} & \multicolumn{1}{c|}{$3$}       & $1$   & $344$ & $458$ & $749$ \\ \midrule
\multicolumn{1}{c|}{Top$3$}                  & \multicolumn{1}{l|}{\texttt{java.util.HashMap\$KeySet}}                                         & \multicolumn{1}{c|}{2}         & $1$   & $104$ & $104$ & $106$ \\ \midrule
\multicolumn{1}{c|}{\multirow{2}{*}{Top $4$}} & \multicolumn{1}{l|}{\texttt{company.core.services.droits}}                                                           & \multicolumn{1}{c|}{21}        & $104$ & $482$ & $541$ & $730$ \\
\multicolumn{1}{c|}{}                        & \multicolumn{1}{l|}{\texttt{company.outils.persistance.IK}}                                                          & \multicolumn{1}{c|}{8}         & $80$  & $82$  & $88$  & $118$ \\  \bottomrule
\end{tabular}
}
 
  \label{tab:factorystats}
\end{table}

\smallbreak
\noindent \textbf{Illustrative results.} We present the top $4$ patterns discovered by \texttt{SCA-MINER} from our dataset in Fig.\ref{fig:toppatterns}. Each pattern is accompanied by a description of its corresponding subgroup and the number of memory snapshots it covers. The red color in the figure represents the expected value of the concept in the antichain, while the green color indicates the average observed values of the subgroup with respect to this concept. We chose to report the mean value in the charts since it is more interpretable, intuitive, and comparable to the expected value. However, in Table\ref{tab:factorystats}, we also provide all the statistics related to the top $4$ patterns, including the minimum and maximum value, and $0.2-$order quantile, which further reveals the properties of the subgroup distributions. Overall, our approach has successfully identified interesting and surprising over-expressed patterns for all the extracted patterns. These patterns are diverse, non-redundant, and cover large subgroups (e.g., $85$ and $183$ memory snapshots in the second and third pattern). Additionally, our approach is capable of outputting generic packages as well as single classes when relevant.

The first pattern discovered by the algorithm \texttt{SCA-MINER} highlights that the server \texttt{PAM-p-03012} experienced a consistent increase in heap space usage for two packages: \texttt{company.outils.persistance} and \texttt{java.sql}. The latter package includes several classes such as \texttt{sql.Timestamp} and \texttt{sql.BigDecimal}, indicating that the memory saturation is due to improper usage of the Direct SQL API in the source code. This API allows direct access to the database from the source code, bypassing the Hibernate (Object-Relational-Mapping) layer. Although it avoids loading Java objects mapped to the database in memory, it can cause memory saturation with SQL objects if not properly handled. The second component of the antichain is the package \texttt{company.outils.persistance}, which covers classes used to identify objects with primary keys, further strengthening the hypothesis of excessive object loading with the Direct SQL API. This hypothesis was confirmed by reviewing past maintenance resolution tickets, some of which occurred shortly after memory saturation. It's important to note that although this incident highlights a memory crash problem, it is not a memory leak, as it occurs rapidly due to an unhandled use case. This confirms that our method is not limited to specific issues and can diagnose any memory-related problem.

The antichain's second pattern is noteworthy because it highlights the class \texttt{LotContQualVal}, which is highly contrastive for all Factory servers, with more than 97 weekly users. This pattern reveals that the frequent use of this particular class in Factory servers tends to cause memory saturation when the number of users exceeds a certain threshold, which is considerably high in the context of an ERP for industries. Despite having the best Information Content (IC), this pattern is ranked second because it has a lower generality and a relatively high depth compared to other patterns (i.e., $\text{DL}_{\text{A}} \text{is larger}$). On the other hand, Despite having a smaller depth, pattern $3$ has been ranked lower than pattern $2$ because it has more selectors ($3$), making it less interpretable than pattern $2$. The Factory servers are crucial because they manage the core of the factory and our clients' production. The \texttt{LotContQualVal} class is instrumental in identifying functionality with an unusual problem, particularly those related to product quality features. This problem leads to a quick saturation event, and many maintenance tickets raised by our clients have identified this class as one of the crucial elements in root-cause analysis.

\begin{figure}
\centering
\subfloat[Size of the classes belonging to \texttt{services.droits} package in pattern 4.]{
\includegraphics[width=0.45\textwidth]{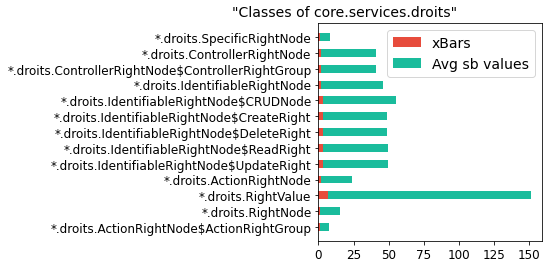}
\label{fig1:subfig1}}
\qquad
\subfloat[Heap consumption of \texttt{services.droits} during 50 days in a leaky server.]{
\includegraphics[width=0.45\textwidth]{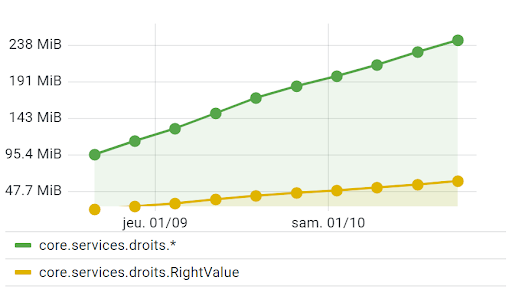}
\label{fig1:subfig3}}
\caption{\label{fig:explication} Explanations of one of  \texttt{SCA-MINER} patterns}
\end{figure}

Pattern 4 highlights a prevalent issue in our ERP, which is memory leaks caused by the \texttt{RightValue} class. Therefore, in this pattern, we provide our experts with a more general solution that identifies memory leak problems across all classes in the \texttt{services.droits} package, rather than just a specific class. This is shown in Fig.\ref{fig1:subfig1}, where we compare the expected value with the subgroup average value for all relevant classes. Through further analysis with our experts, we discovered a growing memory leak that had gone unnoticed for several weeks (Fig.\ref{fig1:subfig3}), which was not detected by other supervision tools. The antichain package helped pinpoint the source of the bug in the source code so that it could be fixed as a temporary solution. However, to prevent this problem from recurring, we established a systematic memory snapshot control to monitor the trend of the package size.

\section{Related work}\label{sec:relatedwork}


\noindent \textbf{Java Memory Analysis.} When it comes to Java memory analysis, identifying classes that leak memory and cause an \texttt{OutOfMemoryError} with a large heap size can be an overwhelming task for developers. One common approach involves analyzing heap dumps with available tools such as \texttt{Eclipse-MAT}\cite{website:eclipsemat} to determine which objects are consuming abnormally large amounts of heap space. However, detecting and diagnosing memory leaks has been the subject of a wide range of related work, including both static and dynamic approaches. Static methods involve formulating leak detection as a reachability problem by identifying value flows from the source \texttt{malloc} to the sink \texttt{free}\cite{DBLP:conf/popl/HackettR05} or detecting static liveness regions~\cite{DBLP:conf/cc/ShahamKS00}. Dynamic methods, on the other hand, rely on either growing types~\cite{DBLP:conf/wosp/WeningerGM19} (i.e., types with a growing number of run-time instances) or object staleness~\cite{DBLP:conf/icse/JungLRP14,DBLP:journals/tosem/XuR13} (i.e., the elapsed time since the last use of an object). However, these methods and/or tools require either the entire heap dumps or the complete code flow diagram as input, which can be prohibitively expensive in industrial production scenarios. Furthermore, unlike our approach, these solutions are applied separately to analyze specific memory problems, rather than considering a generic approach that handles simultaneously a large set of memory incidents.

\noindent \textbf{Subgroup Discovery.} Subgroup Discovery is recognized as an important technique to derive insights or make discoveries from data. 
This well-established data mining task aims to identify interpretable local subgroups that foster some property of interest. A popular case consists in binary target concepts~\cite{DBLP:journals/jmlr/LavracKFT04} that seeks to find subsets of data such that the proportion of a specific target class is significantly higher than expected. A considerable amount of literature on Subgroup Discovery approaches consider complex target attributes such as categorical~\cite{DBLP:conf/pkdd/GrosskreutzRW08}, numerical~\cite{DBLP:journals/datamine/LemmerichAP16,DBLP:journals/datamine/BoleyGGV17} and multi-target~\cite{DBLP:conf/icde/LijffijtKDPOB18,DBLP:journals/datamine/DuivesteijnFK16}. Furthermore, various approaches have been proposed to generalize Subgroup Discovery to complex data structures, e.g., sequential data~\cite{DBLP:conf/pkdd/GrosskreutzLT13} and  graphs~\cite{DBLP:journals/ml/KaytouePZBR17}. Hierarchies had been early taken into account in the pattern syntax \cite{DBLP:conf/kdd/KamberHC97}. However, they remain under-exploited on the target side. None of the existing techniques has exploited this framework for hierarchical target concepts. Our proposed method is based on the \texttt{FORSIED} framework \cite{DBLP:conf/ida/Bie13}, which formalizes the data exploration process as an interactive exchange of information between the model and data analyst, incorporating the analyst's prior belief state. The framework employs Subjective Interestingness (SI) \cite{DBLP:journals/datamine/Bie11}, which has been successfully used to assess subgroups in various structures such as n-ary relations \cite{DBLP:conf/dsaa/LijffijtSKB15} and graphs \cite{DBLP:journals/datamine/DengKLB21,DBLP:journals/datamine/BendimeradMLPRB20}. In \cite{DBLP:conf/kdd/BendimeradLPRB19}, the authors investigated the discovery of subjectively contrastive attributes within a single hierarchy. Our work extends this approach by mining subgroups across multiple hierarchies.

\section{Conclusion and Perspectives\label{ref:conclusion}}

As today's IT environments continue to evolve, growing larger and more intricate, they generate a diverse range of heterogeneous data that necessitates efficient analysis for incident management. This demands effective data-driven approaches rooted in the concept of AIOps, capable of extracting insightful patterns from such data. In this paper, we presented \texttt{SCA-MINER}, the first data mining algorithm enabling the discovery of interesting and contextualized subgroups from data with targets anchored in a hierarchy along with descriptive attributes, while taking into account non-redundancy and interestingness using the SI framework. Thanks to a real-world dataset of Java memory incidents provided by our ERP supervision team, we showcase several actionable patterns, some of which have been seamlessly integrated into our rule-based maintenance engine. Moving forward, we intend to consistently employ the algorithm for the discovery of new and interesting rules, bugs, and memory leaks. Additionally, we plan to extend this methodology to other suitable use cases, such as the analysis of microservices traces that document hierarchy-like call dependencies among microservices. In parallel, we aim to enhance the computational efficiency of the mining algorithm by incorporating and comparing alternative heuristic approaches (e.g., MCTS, genetic algorithms, etc.). For the algorithm's practical usability in daily operations, akin to regular SQL workload analyses conducted in \cite{DBLP:conf/kbse/RemilBMCK21}, particular emphasis should also be placed on the temporal dimension when maintaining prior knowledge.

\bibliographystyle{IEEEtran}
\bibliography{IEEEabrv,references}

\begin{thebibliography}{10}
\providecommand{\url}[1]{#1}
\csname url@samestyle\endcsname
\providecommand{\newblock}{\relax}
\providecommand{\bibinfo}[2]{#2}
\providecommand{\BIBentrySTDinterwordspacing}{\spaceskip=0pt\relax}
\providecommand{\BIBentryALTinterwordstretchfactor}{4}
\providecommand{\BIBentryALTinterwordspacing}{\spaceskip=\fontdimen2\font plus
\BIBentryALTinterwordstretchfactor\fontdimen3\font minus \fontdimen4\font\relax}
\providecommand{\BIBforeignlanguage}[2]{{%
\expandafter\ifx\csname l@#1\endcsname\relax
\typeout{** WARNING: IEEEtran.bst: No hyphenation pattern has been}%
\typeout{** loaded for the language `#1'. Using the pattern for}%
\typeout{** the default language instead.}%
\else
\language=\csname l@#1\endcsname
\fi
#2}}
\providecommand{\BIBdecl}{\relax}
\BIBdecl

\bibitem{dang2019aiops}
Y.~Dang, Q.~Lin, and P.~Huang, ``Aiops: real-world challenges and research innovations,'' in \emph{2019 IEEE/ACM 41st International Conference on Software Engineering: Companion Proceedings (ICSE-Companion)}.\hskip 1em plus 0.5em minus 0.4em\relax IEEE, 2019, pp. 4--5.

\bibitem{bogatinovski2021artificial}
J.~Bogatinovski, S.~Nedelkoski, A.~Acker, F.~Schmidt, T.~Wittkopp, S.~Becker, J.~Cardoso, and O.~Kao, ``Artificial intelligence for it operations (aiops) workshop white paper,'' \emph{arXiv preprint arXiv:2101.06054}, 2021.

\bibitem{becker2020towards}
S.~Becker, F.~Schmidt, A.~Gulenko, A.~Acker, and O.~Kao, ``Towards aiops in edge computing environments,'' in \emph{2020 IEEE International Conference on Big Data (Big Data)}.\hskip 1em plus 0.5em minus 0.4em\relax IEEE, 2020, pp. 3470--3475.

\bibitem{notaro2021survey}
P.~Notaro, J.~Cardoso, and M.~Gerndt, ``A survey of aiops methods for failure management,'' \emph{ACM Transactions on Intelligent Systems and Technology (TIST)}, vol.~12, no.~6, pp. 1--45, 2021.

\bibitem{chen2020towards}
Z.~Chen, Y.~Kang, L.~Li, X.~Zhang, H.~Zhang, H.~Xu, Y.~Zhou, L.~Yang, J.~Sun, Z.~Xu \emph{et~al.}, ``Towards intelligent incident management: why we need it and how we make it,'' in \emph{Proceedings of the 28th ACM Joint Meeting on European Software Engineering Conference and Symposium on the Foundations of Software Engineering}, 2020, pp. 1487--1497.

\bibitem{DBLP:conf/sigsoft/XieA05}
Y.~Xie and A.~Aiken, ``Context- and path-sensitive memory leak detection,'' in \emph{Proceedings of the 10th ESEC-FSE}, M.~Wermelinger and H.~C. Gall, Eds.\hskip 1em plus 0.5em minus 0.4em\relax {ACM}, 2005.

\bibitem{DBLP:conf/icse/JungLRP14}
C.~Jung, S.~Lee, E.~Raman, and S.~Pande, ``Automated memory leak detection for production use,'' in \emph{36th ICSE}, P.~Jalote, L.~C. Briand, and A.~van~der Hoek, Eds.\hskip 1em plus 0.5em minus 0.4em\relax {ACM}, 2014.

\bibitem{DBLP:conf/wosp/WeningerGM19}
M.~Weninger, E.~Gander, and H.~M{\"{o}}ssenb{\"{o}}ck, ``Analyzing data structure growth over time to facilitate memory leak detection,'' in \emph{Proceedings of the 2019 {ACM/SPEC} ICPE}, V.~Apte, A.~D. Marco, M.~Litoiu, and J.~Merseguer, Eds.\hskip 1em plus 0.5em minus 0.4em\relax {ACM}, 2019.

\bibitem{DBLP:journals/jss/SorS14}
V.~Sor and S.~N. Srirama, ``Memory leak detection in java: Taxonomy and classification of approaches,'' \emph{J. Syst. Softw.}, vol.~96, pp. 139--151, 2014.

\bibitem{DBLP:books/mit/fayyadPSU96/Klosgen96}
W.~Kl{\"{o}}sgen, ``Explora: {A} multipattern and multistrategy discovery assistant,'' in \emph{Advances in Knowledge Discovery and Data Mining}.\hskip 1em plus 0.5em minus 0.4em\relax {AAAI/MIT} Press, 1996, pp. 249--271.

\bibitem{DBLP:journals/widm/Atzmueller15}
M.~Atzmueller, ``Subgroup discovery,'' \emph{Wiley Interdiscip. Rev. Data Min. Knowl. Discov.}, vol.~5, no.~1, pp. 35--49, 2015.

\bibitem{DBLP:journals/datamine/Bie11}
T.~De~Bie, ``Maximum entropy models and subjective interestingness,'' \emph{Data Mining and Knowledge Discovery}, vol.~23, no.~3, pp. 407--446, 2011.

\bibitem{DBLP:journals/datamine/KapoorSL21}
S.~Kapoor, D.~K. Saxena, and M.~van Leeuwen, ``Online summarization of dynamic graphs using subjective interestingness for sequential data,'' \emph{Data Min. Knowl. Discov.}, vol.~35, no.~1, pp. 88--126, 2021.

\bibitem{DBLP:journals/datamine/DengKLB21}
J.~Deng, B.~Kang, J.~Lijffijt, and T.~D. Bie, ``Mining explainable local and global subgraph patterns with surprising densities,'' \emph{Data Min. Knowl. Discov.}, vol.~35, no.~1, pp. 321--371, 2021.

\bibitem{DBLP:journals/datamine/BendimeradMLPRB20}
A.~Bendimerad, A.~Mel, J.~Lijffijt, M.~Plantevit, C.~Robardet, and T.~D. Bie, ``Sias-miner: mining subjectively interesting attributed subgraphs,'' \emph{Data Min. Knowl. Discov.}, vol.~34, no.~2, pp. 355--393, 2020.

\bibitem{DBLP:conf/kdd/BendimeradLPRB19}
A.~Bendimerad, J.~Lijffijt, M.~Plantevit, C.~Robardet, and T.~D. Bie, ``Contrastive antichains in hierarchies,'' in \emph{Proceedings of the 25th {ACM} {SIGKDD}}, 2019, pp. 294--304.

\bibitem{DBLP:conf/iccs/GanterK01}
B.~Ganter and S.~O. Kuznetsov, ``Pattern structures and their projections,'' in \emph{{ICCS}}, ser. Lecture Notes in Computer Science, vol. 2120.\hskip 1em plus 0.5em minus 0.4em\relax Springer, 2001, pp. 129--142.

\bibitem{DBLP:books/lib/Bishop07}
C.~M. Bishop, \emph{Pattern Recognition and Machine Learning}, ser. Information Science and Statistics.\hskip 1em plus 0.5em minus 0.4em\relax Springer, 2007.

\bibitem{DBLP:conf/pkdd/Lemmerich018}
F.~Lemmerich and M.~Becker, ``pysubgroup: Easy-to-use subgroup discovery in python,'' in \emph{ECML/PKDD}, vol. 11053.\hskip 1em plus 0.5em minus 0.4em\relax Springer, 2018, pp. 658--662.

\bibitem{DBLP:journals/jmlr/LavracKFT04}
N.~Lavrac, B.~Kavsek, P.~A. Flach, and L.~Todorovski, ``Subgroup discovery with {CN2-SD},'' \emph{JMLR}, vol.~5, pp. 153--188, 2004.

\bibitem{DBLP:journals/datamine/LeeuwenK12}
M.~van Leeuwen and A.~J. Knobbe, ``Diverse subgroup set discovery,'' \emph{Data Min. Knowl. Discov.}, vol.~25, no.~2, pp. 208--242, 2012.

\bibitem{DBLP:conf/flairs/AtzmullerP08}
M.~Atzm{\"{u}}ller and F.~Puppe, ``Semi-automatic refinement and assessment of subgroup patterns,'' in \emph{Proceedings of the 21st AAAI}.\hskip 1em plus 0.5em minus 0.4em\relax {AAAI} Press, 2008, pp. 323--328.

\bibitem{website:eclipsemat}
\BIBentryALTinterwordspacing
eclipse, ``Eclipse memory analyzer (mat).'' [Online]. Available: \url{https://www.eclipse.org/mat/}
\BIBentrySTDinterwordspacing

\bibitem{DBLP:conf/popl/HackettR05}
B.~Hackett and R.~Rugina, ``Region-based shape analysis with tracked locations,'' in \emph{Proceedings of the 32nd {ACM} {SIGPLAN-SIGACT} Symposium on POPL}, J.~Palsberg and M.~Abadi, Eds.\hskip 1em plus 0.5em minus 0.4em\relax {ACM}, 2005, pp. 310--323.

\bibitem{DBLP:conf/cc/ShahamKS00}
R.~Shaham, E.~K. Kolodner, and S.~Sagiv, ``Automatic removal of array memory leaks in java,'' in \emph{Compiler Construction, 9th International Conference, {CC} ETAPS}, ser. Lecture Notes in Computer Science, D.~A. Watt, Ed., vol. 1781.\hskip 1em plus 0.5em minus 0.4em\relax Springer, 2000, pp. 50--66.

\bibitem{DBLP:journals/tosem/XuR13}
G.~Xu and A.~Rountev, ``Precise memory leak detection for java software using container profiling,'' \emph{{ACM} Trans. Softw. Eng. Methodol.}, vol.~22, no.~3, pp. 17:1--17:28, 2013.

\bibitem{DBLP:conf/pkdd/GrosskreutzRW08}
H.~Grosskreutz, S.~R{\"{u}}ping, and S.~Wrobel, ``Tight optimistic estimates for fast subgroup discovery,'' in \emph{{ECML/PKDD} 2008}, vol. 5211.\hskip 1em plus 0.5em minus 0.4em\relax Springer, 2008, pp. 440--456.

\bibitem{DBLP:journals/datamine/LemmerichAP16}
F.~Lemmerich, M.~Atzmueller, and F.~Puppe, ``Fast exhaustive subgroup discovery with numerical target concepts,'' \emph{Data Min. Knowl. Discov.}, vol.~30, no.~3, pp. 711--762, 2016.

\bibitem{DBLP:journals/datamine/BoleyGGV17}
M.~Boley, B.~R. Goldsmith, L.~M. Ghiringhelli, and J.~Vreeken, ``Identifying consistent statements about numerical data with dispersion-corrected subgroup discovery,'' \emph{Data Min. Knowl. Discov.}, vol.~31, no.~5, pp. 1391--1418, 2017.

\bibitem{DBLP:conf/icde/LijffijtKDPOB18}
J.~Lijffijt, B.~Kang, W.~Duivesteijn, K.~Puolam{\"{a}}ki, E.~Oikarinen, and T.~D. Bie, ``Subjectively interesting subgroup discovery on real-valued targets,'' in \emph{34th ICDE}.\hskip 1em plus 0.5em minus 0.4em\relax {IEEE} Computer Society, 2018, pp. 1352--1355.

\bibitem{DBLP:journals/datamine/DuivesteijnFK16}
W.~Duivesteijn, A.~Feelders, and A.~J. Knobbe, ``Exceptional model mining - supervised descriptive local pattern mining with complex target concepts,'' \emph{Data Min. Knowl. Discov.}, vol.~30, no.~1, pp. 47--98, 2016.

\bibitem{DBLP:conf/pkdd/GrosskreutzLT13}
H.~Grosskreutz, B.~Lang, and D.~Trabold, ``A relevance criterion for sequential patterns,'' in \emph{Machine Learning and Knowledge Discovery in Databases - European Conference, {ECML} {PKDD} 2013}, 2013, pp. 369--384.

\bibitem{DBLP:journals/ml/KaytouePZBR17}
M.~Kaytoue, M.~Plantevit, A.~Zimmermann, A.~A. Bendimerad, and C.~Robardet, ``Exceptional contextual subgraph mining,'' \emph{Mach. Learn.}, vol. 106, no.~8, pp. 1171--1211, 2017.

\bibitem{DBLP:conf/kdd/KamberHC97}
M.~Kamber, J.~Han, and J.~Chiang, ``Metarule-guided mining of multi-dimensional association rules using data cubes,'' in \emph{Proceedings of the 3rd International Conference on KDD}, 1997, pp. 207--210.

\bibitem{DBLP:conf/ida/Bie13}
T.~D. Bie, ``Subjective interestingness in exploratory data mining,'' in \emph{Advances in Intelligent Data Analysis {XII} - 12th IDA}, 2013, pp. 19--31.

\bibitem{DBLP:conf/dsaa/LijffijtSKB15}
J.~Lijffijt, E.~Spyropoulou, B.~Kang, and T.~D. Bie, ``P-n-rminer: {A} generic framework for mining interesting structured relational patterns,'' in \emph{2015 {IEEE} International Conference on DSAA}, 2015, pp. 1--10.

\bibitem{DBLP:conf/kbse/RemilBMCK21}
Y.~Remil, A.~Bendimerad, R.~Mathonat, P.~Chaleat, and M.~Kaytoue, ``"what makes my queries slow?": Subgroup discovery for {SQL} workload analysis,'' in \emph{36th {IEEE/ACM} International Conference on ASE}.\hskip 1em plus 0.5em minus 0.4em\relax {IEEE}, 2021, pp. 642--652.

\end{thebibliography}
\end{document}